# Copper Replaces Tin: A Copper based Gelling Catalyst for Poly-Urethane from Discarded Motherboard

*Bibudha Parasar, [§,‡] Gao Wen Jing, [§] Dandan Yuan, [§] Wang Kun, [§] Peng Wang, [§] Arijit Dasgupta, [‡] Atharva Sahasrabudhe, [§,‡] Soumitra Barman, [‡] Rongxin Yuan, [§] Soumyajit Roy* [§,‡].*

[§]School of Chemistry and Materials Engineering, Changshu Institute of Technology, Changshu, Jiangsu, P. R. China.

[‡]Eco-friendly Applied Materials Laboratory (EFAML), DCS, New Campus, Indian Institute of Science Education and Research, Kolkata, India.

ABSTRACT. A discarded motherboard based eco-friendly copper catalyst has been programmed to replace the industrially used tin based catalyst DBTDL. The catalyst has been characterized by UV-Vis spectroscopy, FT-IR and TEM. Using the catalyst reaction conditions is optimized and under the optimized condition, both polyurethane and polyurethane foam are prepared, thus proving the generality of the catalyst to be used in industries. A possible mechanism has also been proposed.

**Introduction:**

Over the past decades, polyurethane (PU) has gained immense importance in contemporary and futuristic material science. It has become important as elastomers, thermoplastics, thermorigids, adhesives, coatings, sealants, fibers [1]. Hence demand for PU catalyst has also increased steadily. On the other hand, discarded computer hardware poses an imminent environmental threat. Here we synthesize PU catalyst using parts of discarded computer motherboard. It is hence an attempt to solve two problems simultaneously. This approach transforms the discarded computer motherboard to a resource for catalyst to meet the need of a material (PU) in high demand. Before we proceed further, we shortly review the PU catalyst in use as of today.

Hard polyurethane catalysis needs gelling catalyst whereas polyurethane foam catalysis requires both gelling catalyst and blowing catalyst serving different roles. A gelling catalyst favours the reaction between isocyanate and alcohol whereas a blowing catalyst facilitates the reaction between isocyanate and water. Moreover a desired PU catalyst would also require cyclo-trimerization catalyst to harden the material, oxazolidone catalysts and carbodiimide catalysts to impart thermal resistant properties in addition to gelling and blowing actions [2].

Although several tertiary amine catalyst such as DABCO [3], 1,8-Diazabicyclo,5,4,0,undecene-7 [4], 2-Methyl-2 azanorbornane [5] have been used as blowing catalyst and several delayed action catalysts such as 1,8-Diazabicyclo (5,4,0) undecene (7DBUTM) [6] Polycat SA-1TM, Polycat SAS-102TM, Polycat SA-610/50TM [4,6] have been used to control reactivity profiles, they all have a problem. Amine catalysts remaining in the foam structure evaporate and often release an amine odor giving rise to fogging of foam [2]. So, alternative catalysts have been used as blowing agents [7]. Moreover non-emission catalysts such as tertiary amines with reactive groups such as -OH or -$NH_2$ have also been used. Triethylendiamine based catalyst [8] and commercially available amine N-substituted morpholine [9] are often used to avoid fogging and odour problems. Even tertiary alkanolamine based catalysts are used for PU foam preparation [10]. A dendrimeric catalyst based on Methylated poly(propylene imine) has been developed to prepare odorless PU foam [11]. Several other amine based catalyst such as N,N-dimethylalkylamines [12], N-alkylpyrrolidines[13], bis(dimethylaminopropyl)amine [14], N-substituted perhydrodioxazepines [15] have also been developed. Moreover commercially available compounds such as guanidine [16], thiourea [16], urea derivatives [17,19] have also been explored in PU catalysis. To tackle tackiness, another problem of PU, an aminoimide [18] based catalyst, which could give desirable tack-free times from components of polyols and polyisocyanates has been used. Some other catalysts such as quaternary hydroxyalkyl tertiary amine[20,21], aminoborate esters [22], aminioorthesters [23], diimines [24] have also been developed. Green alternatives like, ionic liquid based catalyst with different cation components such as imidazolium [26], phosphonium [27], iminium [28], quaternarized borates and non-fugitive catalysts have also been used for the preparation of polyurethane [25-31]. Although concerted efforts exist to phase out amine catalysts to rid the foam of fogging problem, little has been done to replace the traditional gelling catalyst DBTDL (dibutiltin (IV) dilaurate) [32]. For instance, acetylacetone complexes of different transition metals are chemically synthesized and used as catalyst. However several other compounds such as stannous 2-ethylhexanoate, stannous octanoate, dibutyltin 2-ethylhexanoate, dibutyltin diacetate, dioctyltin dimercaptide [33] are used as gelling catalyst; they all have toxicity and price issues. Hence there is a need for an eco-friendly catalyst that is cheap, efficient and effective. We now describe the design of our catalyst.

Since gelling catalyst serves as the backbone of polyurethane catalysis [2], we felt, the need to experiment to replace the traditional tin based catalyst. So, we ask, is it possible to replace the tin based conventional catalyst by some non-toxic eco-friendly catalyst? Looking at the pre-existing catalyst, for our catalyst design, we conclude that, to catalyze the polyurethane reaction between a di/polyol and di/polyisocyanate, the catalyst should fulfill the following requirements. 1) It must have long hydrophobic chain, as they tend to have a hydrophobic environment to stabilize the emerging oligomers. 2) It should have a metal component (to hold the Lewis acid character) source that is cheap and 3) it is easily available. In recent years, copper has emerged as a non-toxic versatile catalyst for diverse transformations and is abundant in computer motherboard [34-48]. We hypothesized: it is perhaps possible to design a catalyst with copper as the central metal atom, with oligomers stabilizing long chain as its counter ion. Copper is also easily extractable from discarded computer motherboard. Our intrinsic hypothesis is based on the assumption that, both the alcohol and isocyanate chelates the central metal atom copper in the first step. Subsequently the activated alcoholic hydrogen then transfers its hydrogen to the isocyanate bonded to the copper in a fashion similar to the mechanism to that of DBTDL [49]. The motherboard based copper complex finally releases the urethane molecule to regenerate the catalyst.

To test this hypothesis, copper is extracted from discarded computer motherboard. As mentioned earlier, this choice is prompted by the growing menace of discarded computer motherboard in enhanced heavy metal poisoning especially in emerging economies. Toxic metal scrapes from computer hardware are sometimes manually scavenged or left unattended to enter into ecosystem causing severe environmental problems. Hence our choice to utilize different copper components of a discarded motherboard as catalyst would address the environmental pollution issues. Bamboo is chosen as the source of polyol. Commercially available IPDI is used as an isocyanate. DBSNa (Sodium dodecylbenzenesulfonate) is chosen as the surfactant counter ion as an interface agent for copper ions. Using this motherboard based copper catalyst we have prepared hard polyurethane and polyurethane foam.

**Experimental Section**

Polyurethane is prepared in a sequential manner starting from preparation of the catalyst to its synthesis. The entire sequence is schematically summarized in Figure 1.

*Preparation of bamboo polyol:*

Bamboo polyol is prepared according to a previously reported procedure [50]. In a typical method, to a 500mL two necked round bottom flask equipped with a reflux condenser, 40g of PEG-400 and 10 g of glycerol is added. The resulting solution is stirred for 10 minutes at room temperature followed by acidification with 1.5g of concentrated Sulfuric acid. 20g of bamboo (Phyllostachys pubescens) residue, in the form of finely cut pieces within the size range of 0.1cm to 0.7 cm, is added to this mixture. The reaction mixture is maintained at a temperature of 160 ºC for 2 hours with vigorous stirring. Completion of the reaction is indicated by a colour transformation when the initial colorless, inhomogeneous solution turns black. The reaction is quenched by dipping the flask in ice-cold water. Subsequently, the contents of the flask are transferred to a sealed polypropylene container and stored at 10 ºC in a refrigerator. Acid and hydroxyl value of the so formed bamboo polyol are measured for subsequent experimentations.

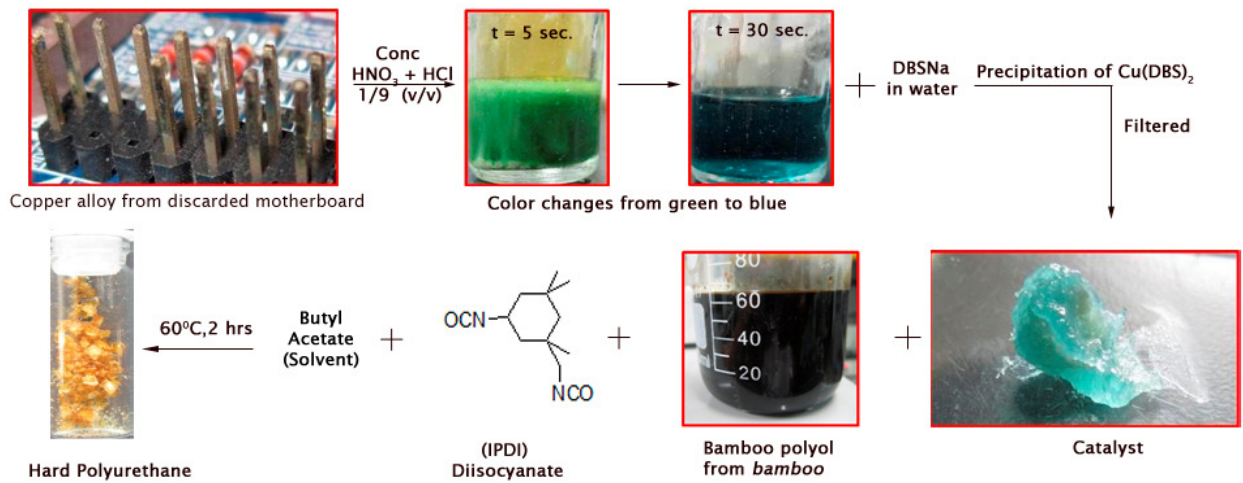

**Figure 1.** Sequential pathway for preparation of polyurethane.

*Acid values of the bamboo polyol:*

Acid value of the prepared bamboo polyol is calculated according to a previously reported procedure and is subsequently used for determination of hydroxyl value [51]. 1 g bamboo-polyol sample is first mixed with 20 mL dioxane-water solution (4/1, v/v). The resulting mixture is titrated against 1 M NaOH till the pH reaches pH 8.3 to indicate the end-point. Acid value is determined using the following expression:

$$Acid\ Value = (V_{NaOH(sample)} - V_{NaOH(blank)}) \times N_{NaOH} \times 56.1/W_{sample}$$

Where,

$V_{NaOH(sample)}$ = volume of NaOH standard solution consumed in sample titration (mL)

$V_{NaOH(blank)}$ = volume of NaOH standard solution consumed in blank titration (mL)

$W_{sample}$ = sample weight (g)

$N_{NaOH}$ = equivalent concentration of NaOH standard solution (M)

*Hydroxyl value of bamboo polyol:*

Phthalic anhydride solution is prepared by dissolving 150 g phathalic anhydride in a 1 L mixture of dioxane and pyridine (9/1, v/v). 10 mL of the phthalic anhydride solution is mixed with 1 g of polyol sample and the final mixture is added into a 150 mL beaker covered with aluminum foil. The beaker is put into a boiling water bath for 20 min. After cooling it down, 20 mL of dioxane-water solution (4/1, v/v) and 5 mL of water are added to the beaker. The resulting mixture is titrated against 1 M NaOH till the pH reaches pH 8.3 to indicate the end-point. Blank titration is conducted using the same procedure. The hydroxyl value is calculated using the following expression:

$$Hydroxyl\ value = (V_{NaOH(blank)} - V_{NaOH(sample)}) \times N_{NaOH} \times 56.1/W_{sample} + Acid\ value$$

Where, the symbols carry their usual meaning as mentioned above.

*Preparation of the catalyst:*

0.4 g of copper-alloy is removed from a discarded motherboard and dissolved in 3 mL of concentrated $HNO_3$. A 0.095 M aqueous solution of sodium dodecylbenzenesulfonate (DBSNa) is prepared by dissolving 2 g of DBSNa in 60 mL of water. The copper solution is added drop wise to aqueous DBSNa with mild stirring. The so formed blue, turbid solution is allowed to settle for 10 minutes. The solid product is collected by filtration and is dried at room temperature for 2 days prior to further use.

*Preparation of hard polyurethane:*

To a neat glass vial, 0.5 g of bamboo polyol and 0.5 g of IPDI is added and the mixture is stirred well for 2 minutes. 0.1 g of the above prepared catalyst is weighed and dissolved in 1 mL of butyl acetate and mechanically mixed for 5 minutes till all the catalyst is solubilized. 300 µL of this solution is taken in a clean pipette and added to the reaction mixture. The solution is stirred well for 10 minutes at room temperature to allow them to mix. The reaction temperature is then gradually increased to 60 ºC and the temperature is maintained for 2 hours. After the completion of the reaction, the hardened polyurethane is collected and washed with (acetic acid+water) mixture in a ratio of 1:20. After washing three times, it is air-dried prior to any further characterization.

*Preparation of polyurethane foam:*

To a neat glass vial, 0.5 g of bamboo polyol and 0.5 g of IPDI is added and the mixture is stirred well for 2 minutes. 0.1 g of the above prepared catalyst is weighed and dispersed in 1 mL of water and mechanically mixed for 5 minutes till it creates a uniform milky white dispersion. 300 µL of this solution is taken in a clean pipette and added to the reaction mixture. The solution is stirred vigorously for 30 minutes at room temperature. The foam is allowed to rise at room temperature. After that the foam is cured at room temperature for 4 days. The polyurethane foam is collected and washed very slowly with (acetic acid+water) mixture in a ratio of 1:20. After washing three times, it is air-dried prior to any further characterization.

*Spectroscopy and microscopy experiments:*

The Fourier transform Infrared (FTIR) spectra reported in this study are recorded as KBr pellets with a Nicolet 380 FT-IR spectrometer in the range 4000-500 $cm^{-1}$. Optical images are taken by a Zeiss axiocam camera. SEM images are acquired by Hitachi table-top Microscope,

Model No. TM3000. TEM images are taken by a Jeol JEM 2010 electron microscope operated at 300 kV.

**Results and discussion:**

*Optimization of the catalyst preparation:*

To prepare an effective catalyst, it is important to first dissolve the motherboard based copper interfaces in a suitable liquefying medium. To do so, the interfaces are removed from the motherboard interfaces, and are dissolved in concentrated acid medium. Concentrated acids of HCl, $HNO_3$, $H_2SO_4$, $CH_3COOH$ and $H_3PO_4$ are screened as the solubilizing medium. In all the experiments 0.16 g of copper is dissolved in 1 mL of concentrated acid. Out of all the acids only $HNO_3$ is effective and dissolves the computer based motherboard interfaces in 30 seconds (Table 1, entry 1). In the process, the solution undergoes a transition in appearance from greenish yellow to light blue (Fig. 1). In HCl, the process is very slow and in other acids, no considerable progress of the reaction is monitored. With the aim to reduce time further so as to reduce the total cost of polyurethane, two acid systems are screened. To our surprise, 10 vol% of HCl in $HNO_3$ dissolves the copper in 10 seconds (Table 1, entry 2). Further reduction of time is attempted by increasing the volume percentage of HCl, but it takes longer time and the trend goes on increasing (Table 1, entry 3-10) (Fig. 2).

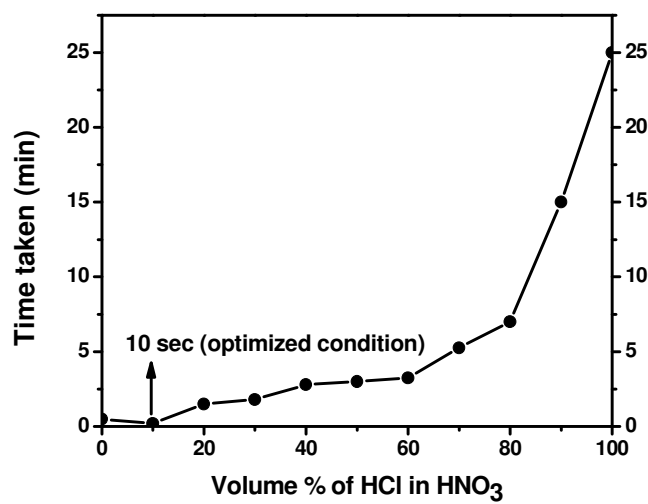

**Figure 2**. Time taken for the copper interfaces to dissolve as a function of volume percentage of HCl in $HNO_3$.

So, 10 vol% of HCl in $HNO_3$ is chosen as the optimized solvent system for digesting copper from the motherboard of the catalyst. Owing to the change in colour of the solution (from green to blue) and due to the existing oxidizing environment, it is reasonable to think that, copper is in cupric (II) state.

**Table 1.** Optimization of the liquefying system for copper interfaces

| Entry | Volume % of HCl in $HNO_3$ | Time |
|---|---|---|
| 1 | 0 | 30 sec |
| 2 | 10 | 10 sec |
| 3 | 20 | 1 min 30 sec |
| 4 | 30 | 1 min 50 sec |
| 5 | 40 | 2 min 50 sec |
| 6 | 50 | 3 min |
| 7 | 60 | 3 min 15 sec |
| 8 | 70 | 5 min 10 sec |
| 9 | 80 | 7 min |
| 10 | 90 | 15 min |
| 11 | 100 | 25 min |

It is hypothesized that, the blue color of the solution might be because of formation of copper nitrate in the solution as addition of ammonium hydroxide to a 20 fold diluted copper solution results in appearance of intense blue precipitate and further addition of excess ammonium hydroxide results in a intense blue solution, which is attributed to first formation of $Cu(OH)_2$ and subsequent formation of $[Cu(NH_3)_4]^{2+}$ upon ammoniacal treatment. Also addition of sodium sulfide solution results in immediate formation of dirty-black precipitate thereby confirms the prime component to be copper (in cupric state) in the solution and hence in the catalyst (Fig. 3)

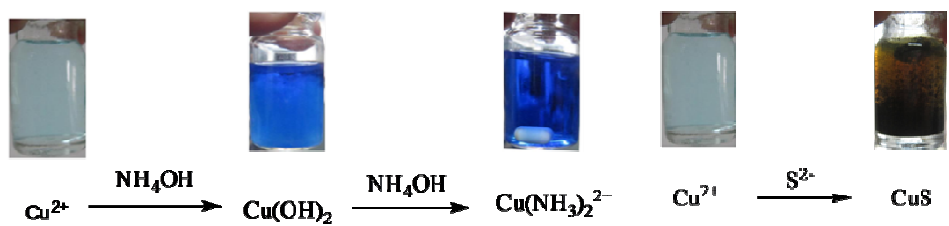

**Figure 3.** (Left) Stepwise addition of $NH_4OH$ to $Cu^{2+}$ solution; (Right) Formation of black CuS precipitate.

To further follow-up our working hypothesis, the prepared copper solution is added to a solution of anionic surfactant to get a copper catalyst with long charged hydrophobic chains as counter ion. Commonly used surfactant sodium dodecyl sulfate (SDS) is chosen as the starting point of our investigation; however it does not lead to compartmentalization. So, other anionic surfactant molecule such as sodium laurate, sodium palmate and sodium stearate are also tried. But they also fail to give any isolable solid product, however initial turbidity are observed. We believe precipitation does not occur possibly because of lack of hydrophobicity. So, it is envisaged, use of a long hydrophobic chain with an aromatic ring may result in enhanced hydrophobicity, where the π-π stacking between two molecules direct the compartmentalization of the solvent system and helps in precipitation of the catalyst. Sodium salt of dodecylbenzenesulfonate (DBSNa) results in absolute compartmentalization of the biphasic system, which is subsequently filtered, air dried and used in reaction protocol optimization. We propose stoichiometry: in a typical experiment 0.65 g of copper results in 6.91 g of catalyst, hence the molar ratio of copper to the counter ion is very close to 1:2. Each $Cu^{2+}$ has two $DBS^-$ and condition the likely composition of the catalyst is $Cu(DBS)_2$, which is evidenced from the stoichiometrically achieved catalyst.

In summary, the catalyst is finally prepared by adding the blue solution to a solution of DBSNa (Sodium Dodecylbenzenesulfonate) in water. Formation of light blue turbidity is attributed to the displacement of chloride and nitrate by DBS (dodecylbenzenesulfonate), which results in the interaction of hydrophobic chain of DBS with water. The origin of turbidity is attributed to compartmentalization of the continuous solvent system (water) by the hydrophobic chains of DBS with water. Such compartments are blue turbid compartments rich in the $Cu(DBS)_2$ species, efficient for catalysis. The solution is then filtered off slowly to give a blue oily solid-paste, containing the catalyst.

*Optimization of the reaction condition:*

Taking the hydroxyl value for crude glycerol as 991mgKOH/g and that for PEG 400 as 268mgKOH/g, hydroxyl value for the bamboo polyol comes out to be 186 mgKOH/g, which is very much close to the value reported previously by others [50]. The model reaction between bamboo polyol and IPDI **1** in the presence of the discarded motherboard based Cu(DBS)$_2$ catalyst is chosen as the starting point of our investigation (Table 2, entry 1) (Scheme -1). The reaction mixture is heated at 60 °C and after 7 hours a solid product is formed. The IR of the solid sample is compared with the DBTDL based polyurethane and FT-IR spectra confirmed it to be polyurethane (Fig. 6 c,d). With the aim of further reducing the reaction time, subsequent optimization is done with the fine-tuning of the reaction parameters (Table 2, entry 2-14) (Fig 4).

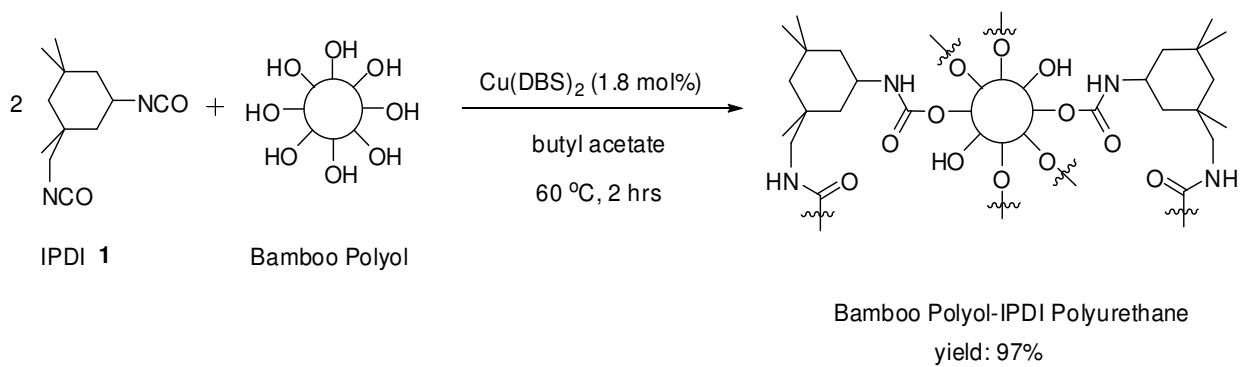

**Scheme 1.** Polyurethane from bamboo polyol and IPDI.

Initial optimization is done with respect to the reaction temperature with 1.8 mol% of catalyst. At 60 °C, polyurethane is afforded in good yield after 2 hours (97%, Table 2, entry 2). When the temperature is increased to 80 °C and beyond that, the yield gradually decreases possibly due to partial decomposition of polyurethane at higher temperature (Table 2, entry 3-5) (Fig. 4a). When the reaction is tried at lower temperature, the yield decreases (Table 2, entry 6). We further try to optimize the reaction temperature with the catalyst loading in order to optimize reaction temperature. At 50 °C when the catalyst loading is increased to 3 mol% and beyond, polyurethane is although afforded with high yield; the time taken for the reaction to complete is 8 hours (Table 2, entry 7-8). Since relatively high yield (Table 2, entry 10-14) is obtained at various loadings at 60 °C, the reaction temperature was fixed at 60 °C.

**Table 2.** Optimization of the reaction condition

| Entry | Catalyst (mol %) | Time (hour) | Temperature ($^0$C) | % Yield |
| --- | --- | --- | --- | --- |
| 1 | 0.3 | 7 | 60 | 83 |
| 2 | 1.8 | 2 | 60 | 97 |
| 3 | 1.8 | 2 | 80 | 94 |
| 4 | 1.8 | 2 | 100 | 92 |
| 5 | 1.8 | 2 | 120 | 90 |
| 6 | 1.8 | 8 | 50 | 83 |
| 7 | 3.0 | 8 | 50 | 97 |
| 8 | 6.0 | 8 | 50 | 95 |
| 9 | 0.0 | 7 | 60 | 36 |
| 10 | 0.6 | 7 | 60 | 95 |
| 11 | 0.9 | 4.5 | 60 | 97 |
| 12 | 1.2 | 4 | 60 | 98 |
| 13 | 3.0 | 2 | 60 | 96 |
| 14 | 6.0 | 2 | 60 | 97 |

In the next attempt to maximize the yield with respect to the catalyst loading, an increase in yield is observed with an increase in catalyst loading as expected (Table 2, entry 1,2,9-11) (Fig. 4b). At 0.9 mol% of catalyst loading, PU is afforded in quantitative yield. Our attempt to decrease the time required to complete the reaction with increasing the catalyst loading fail as it does not result in substantial decrease in reaction time. For instance, even at 6 mol% catalyst loading, it takes two hours for the reaction to go to completion. So, simultaneous optimization of the three axes (time, temperature and catalyst loading) finally leads to the optimized reaction protocol (Table 2, entry 2), which furnishes nearly full conversion of reactants to give the desired polyurethane (Fig. 5).

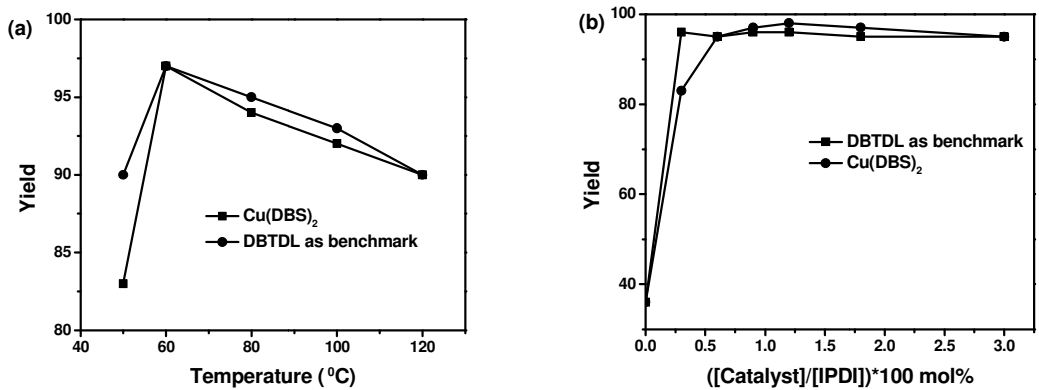

**Figure 4.** a) Temperature dependent polyurethane yield (Catalyst). b) Effect of catalyst loading on polyurethane yield.

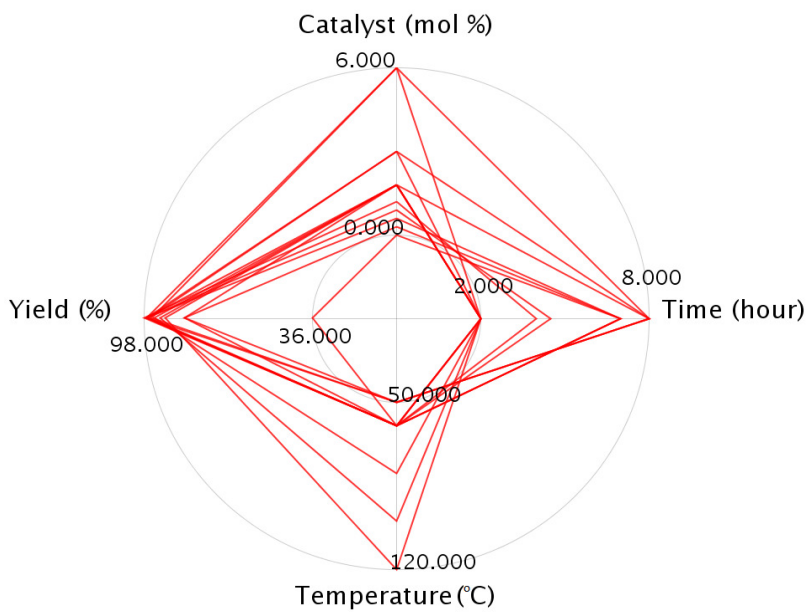

**Figure 5.** Spider plot of the optimization process.

*Efficiency of the catalyst*

The results from the prepared catalyst are compared with respect to that of the conventional catalyst DBTDL. DBTDL is used as a benchmark catalyst here. It turns out, at 0.3 mol% loading of DBTDL, polyurethane is obtained in 95% yield, within 30 minutes (Fig. 4b), whereas 0.6 mol% of loading Cu(DBS)$_2$ catalyst affords PU in 95% yield (Table 2, entry 10). The polyurethane so synthesized is characterized with IR, SEM and optical stereo microscope. The catalyst is tested for its substrate scope and its generality. When methylene diphenyl diisocyanate (MDI) **2** is used, PU was obtained in 97% yield.

Transmission electron microscope (TEM) and high resolution TEM images of the catalysts were taken in butyl acetate (Fig. 6). TEM images shows agglomerated irregular network suggesting weak interaction driven self assembly of the catalyst system.

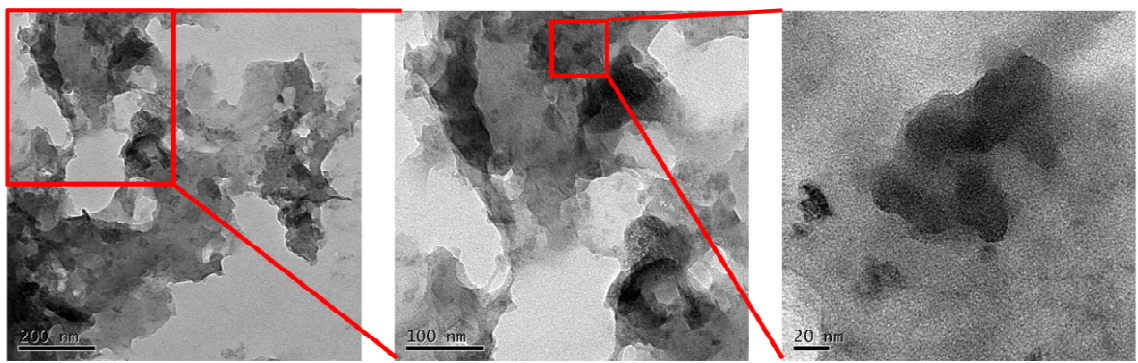

**Figure 6.** Transmission electron microscopy images of the motherboard based copper catalyst.

Encouraged by our success in synthesizing polyurethane, it is checked if the motherboard based copper catalyst can catalyze the PU foam reaction at room temperature. It is worth noting that, at 0.6 mol% catalyst loading PU foam is obtained at nearly quantitative yield, thus proving its general applicability as a PU gelling catalyst. The foam is characterized with IR, and optical stereo microscope. The substrate scope is also checked for foam with MDI.

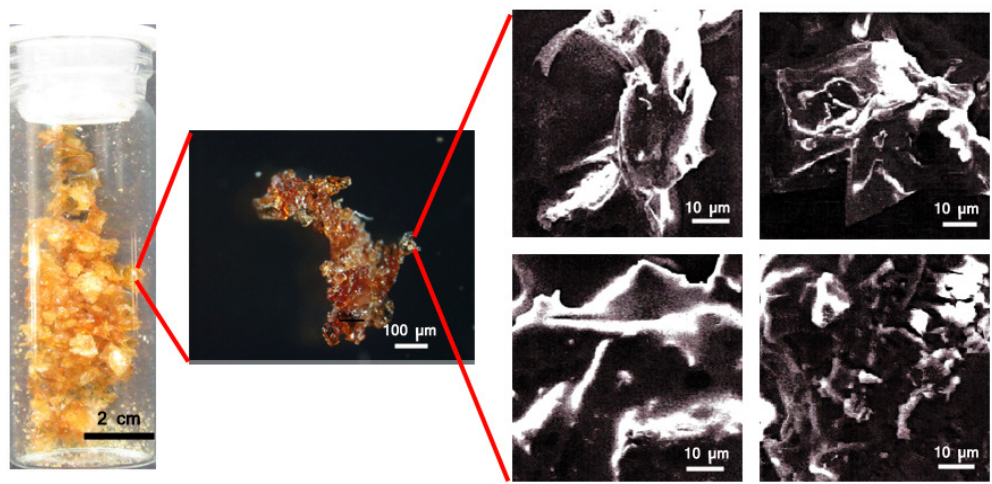

**Figure 7.** Dark field stereo pictures and SEM images of hard polyurethane with IPDI**.**

The dark field optical stereo images are taken, which clearly shows the compactness of the material, which in turn implies at efficient formation of polyurethane (Fig. 7). SEM images also show similar morphology (Fig. 7). Optical images of the polyurethane foam are also taken, which show pores with sizes ranging from 100 µm to 300 µm (Fig. 8).

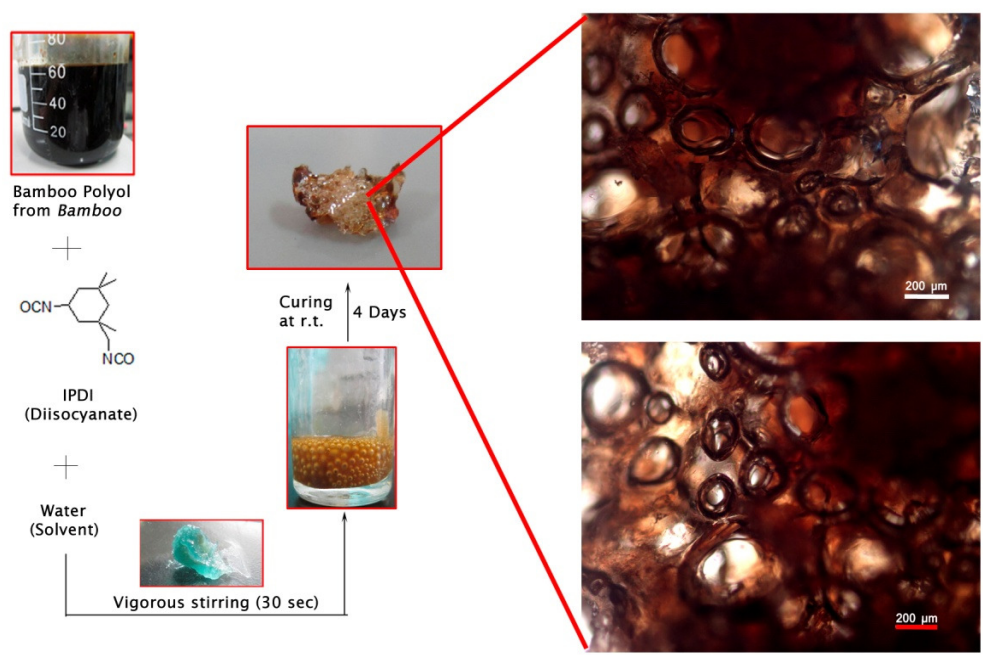

**Figure 8.** Synthetic scheme and optical stereo images of polyurethane foam.

However, open question remains: why does the motherboard based copper catalyst take 2 hours to complete the reaction even at 6 mol% catalyst loading, where as a small amount of DBTDL can catalyze the reaction within 30 minutes? The results completely fit into our hypothesis. In the DBTDL based mechanism, in the last step the laurate acts as the nucleophile and displaces the urethane product. On the contrary, in this copper based mechanism, there is no such nucleophile to displace urethane. So, the neutral molecule urethane undergoes ligand elimination to regenerate the catalyst, which is possibly a slow equilibrium process. Current investigations are on the way to understand the mechanism.

*Spectroscopic Characterization:*

Fourier Transform Infrared spectroscopy is employed to analyze the catalyst (Fig. 9). The catalyst exhibits the characteristics of the fundamental and the split $\nu_3$(S-O) bond vibration. The two prominent vibration of $SO_3$, i.e. $\nu_{as}(SO_3^-)$ and $\nu_s(SO_3^-)$ are observed at 1192cm$^{-1}$ and 1041cm$^{-1}$ respectively, thus proving the existence of dodecylbenzenesulfonate as the counter ion for copper. The complex between copper and the counter ion can be easily realized by comparing the stretches in the sulfonate region. The newly formed copper oxygen bond decreases the electron density on oxygen, thereby inducing a slight red shift in the corresponding ν(SO$_3^-$) vibration frequency. The peak at 3225cm$^{-1}$ on the other hand is attributed to the hydrogen bonding between the proton and the sulfonic oxygens of two molecules, where the proton mediates the stacking of aromatic regions of catalyst. Such a situation is also reflected in peak broadening of the sulfonate region. However, we cannot rule out the possibility of self-stacking of free sulfonate ions above the copper sulfonate complex.

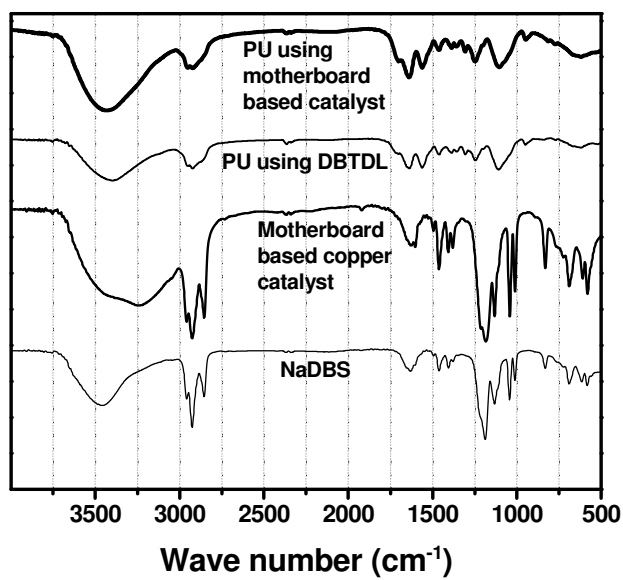

**Figure 9.** IR of a) Sodium dodecylbenzenesulfonate, b) Copper based catalyst, c) Polyurethane using DBTDL, d) Polyurethane using the copper based catalyst.

A typical FT-IR spectrum of the polyol-IPDI polyurethane prepared at the optimized condition is shown (Fig. 10 a). The peak at the wave-number 3436 cm$^{-1}$ is attributed to the stretching vibration of N-H: ν(NH). The peak at 2921 cm$^{-1}$ is caused by the asymmetric stretching vibration of CH$_2$ or ν$_{as}$(CH$_2$). A distinct characteristic peak at 1640 cm$^{-1}$ is attributed to the abundance of amide I bands, which clearly shows the formation of polyurethane. A strong stretching C-O-C vibration at 1100 cm$^{-1}$ proves the participation of both polyol and diisocyanate and hence corroborates the abundance of allophanates. The clear peaks at 1545 cm$^{-1}$ is assigned to the amide band II [δ(N-H) + ν(C-N)]. Evidence of PU formation even comes from the vibration at 1230 cm$^{-1}$, which is attributed to the amide III band [ν(C-N) + δ(N-H)]. The IR of PU foam also shows similar vibration frequency (Fig. 10).

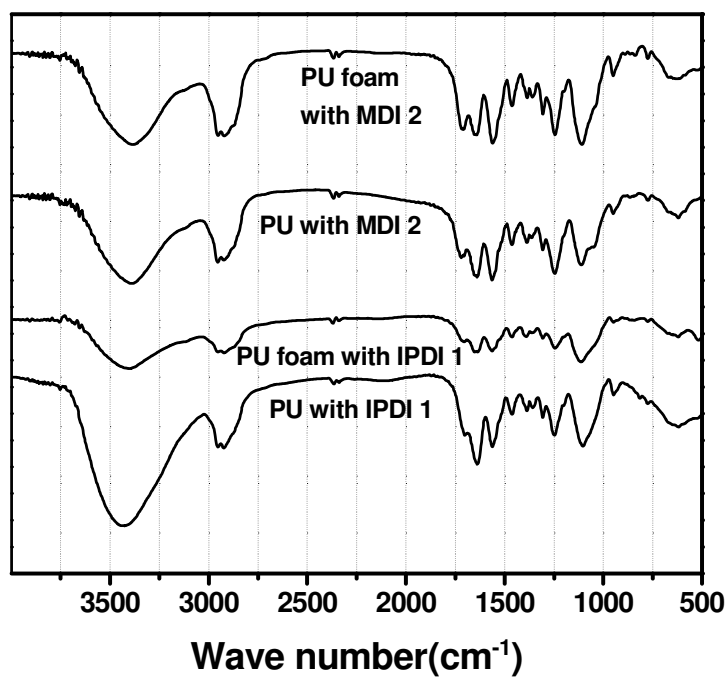

**Figure 10.** IR of the a) PU with IPDI **1**, b) PU foam with IPDI **1**, c) PU with MDI **2**, d) PU foam with MDI **2.**

**Conclusion**

In summary, we have developed a green protocol for the polyurethane reaction by a discarded motherboard based copper catalyst. The copper-based catalyst reported here is non-toxic and does not require any synthetic modification and under the optimized condition with the three parameters (time, temperature and catalyst loading), it could promote the polyurethane reaction in nearly quantitative yield within 2 hour at 60 °C. Furthermore, synthesis of polyurethane foam proves the generality of the catalyst and its potential to displace the currently used toxic and expensive catalysts. The next step would be in the direction of a complete green protocol for the synthesis of polyurethane with starting materials (di/polyol and di/polyisocyanide) being derived from a single environmentally friendly preferably a discarded source. A discarded resource for PU catalyzed by a discarded motherboard catalyst would be the next step forward for a greener tomorrow.

.

AUTHOR INFORMATION

**Corresponding Author**

*E-mail: s.roy@iiserkol.ac.in.

**Author Contributions**

All authors have given approval to the final version of the manuscript.

**Funding Sources**

SR gratefully acknowledge the start-up grants from CIT, P. R. China and IISER-Kolkata, India.

REFERENCES

1. Dieterich, D; Schmelzer, H.G. In *Polyurethane Handbook,* Oertel, G., Ed.; Hanser Publishers: Munich, 1994; 29.

2. Ashida, K. In *Polyurethane and Related Foams: Chemistry and Technology*, CRC Perss, Taylor & Francis Group, 2007.

3. Orchin, M. (Houdry Process Corp.). U.S. Patent 2939851, 1960.

4. Hashimoto, S. In *International Progress in Urethanes*, 3, K. Ashida, K. C. Frisch, , Eds., Technomic: Lancaster, PA, 1981; page 43

5. Hubert, H.H; Gluzek, K.H. In *Proceedings of the FSK/SPI Polyurethane World Congress* Aachen, F.R. Germany, Septmember 29-October 2, 1987; page 820.

6. Casati, F.M.N; Arbier, F.W. In *Proceedings of the SPI 27th Annual Technical/Marketing Conference*, October 20–22, 1982, Bal Harbour, FL, page 35.

7. Reed, D. In *Urethane Technology*, 2000, Vol. 17 June/July; page 22.

8. Cauwenberge, G.V.V; Melder, J.P; Anders, J.T; Benisch, C; Klopsch, R; Daun, G; Dully, C; Buschhaus, B; Boeckemeier, H; Pox, E. PCT Int. Appl., WO 2009030649 A2 20090312, 2009

9. Strachota, A; Strachotova, B; Spirkova, M. In *Materials and Manufacturing Processes*, 2008; Vol 23(6); page 566-570.

10. Burdeniuc, J.J. Eur. Pat. Appl. EP 1457507 A2 20040915, 2004.

11. Froehling, P.E; Corstjens, T. In *Polymeric Materials Science and Engineering*, 1997; vol 77; page 534-535.

12. Jachimowicz, F. U.S. US 4563484 A 19860107, 1986

13. Jachimowicz, F. U.S. US 4450246 A 19840522, 1984

14. McEntire, E.E. Can. CA 1132292 A2 19820921, 1982

15. Zimmerman, R.L; McEntire, E.E. U.S. US 4342687 A 19820803, 1982

16. Horacek, H; Barl, M; Wurmb, R; Marx, M. Ger. Offen. DE 3003963 A1 19810813, 1981

17. Zimmerman, R.L. U.S. US 4242467 A 19801230, 1980

18. Zimmerman, R.L. U.S. US 4156760 A 19790529, 1979

19. McEntire, E.E. U.S. US 4094827 A 19780613, 1978

20. Bechara, I.S; Carroll, F.P; Mascioli, R.L; Panchak, J.R. U. S. Publ. Pat. Appl. B US 497194 I5 19760203, 1976.

21. Bechara, I.S; Carroll, F.P; Mascioli, R.L. U.S. US 3892687 A 19750701, 1975.

22. Bechara, I.S; Holland, D.G. U.S. US 3853818 A 19741210, 1974.

23. Bechara, I.S; Holland, D.G. U.S. US 3786029 A 19740115, 1974.

24. Brizgys, B. Ger. Offen. DE 1931665 B2 19760102, 1970.

25. Neu, O; Siemer, M; Altenhoff, A.G; Schaefer, H; Steinbrecher, A.M. PCT Int. Appl. WO 2011061314 A1 20110526, 2011

26. Athey, P; Wilmot, N; Keaton, R; Babb, D; Boyer, C; Morley, T. PCT Int. Appl. WO 2010054317 A2 20100514, 2010

27. Athey, P; Wilmot, N; Keaton, R; Boyer, C; Morley, T. PCT Int. Appl. WO 2010054313 A2 20100514, 2010.

28. Athey, P; Wilmot, N; Keaton, R; Boyer, C; Morley, T. PCT Int. Appl. WO 2010054311 A2 20100514, 2010.

29. Richmond, J.M; White, K.B. PCT Int. Appl. WO 8303415 A1 19831013, 1983.

30. Wendel, S.H; Mercando, L.A; Tobias, J.D. UTECH 2000, Conference Paper Abstracts, The Hague, Netherlands, Mar. 28-30, 2000 (2000) I6/1-I6/10

31. Mercando, L.A; Kniss, J.G; Tobias, J.D; Plana, A; Listemann, M.L; Wendel, S. In *Polyurethanes Expo'99, Proceedings of the Polyurethanes Expo'99*, Orlando, FL, United States, Sept. 12-15 1999; page 103-134.

32. Lima, V; Pelissoli, N.S; Dullius, J; Ligabue, R; Einloft, S; *Journal of Applied Polymer Science*, **2010**, *115*, *1797–1802*

33. Ashida, K; Hata, K. (Nisshinbo Ind.) Japanese Examined Patent Publication Sho-37-17398, 1962.

34. Reymond, S; Cossy, J. *Chem. Rev.* **2008**, *108 (12)*, 5359–5406.

35. Stanley, L.M; Sibi, M.P. *Chem. Rev.*, **2008**, *108 (8)*, 2887–2902.

36. Poulsen, T.B; Jørgensen,K.A. *Chem. Rev.*, **2008**, *108 (8)*, 2903–2915.

37. Alexakis, A; Bäckvall, J.E; Krause, N; Pàmies, O; Diéguez, M. *Chem. Rev.*, **2008***, 108 (8)*, 2796–2823

38. Chemler, S.R; Fuller, P.H. *Chem. Soc. Rev.*, **2007**, *36,* 1153-1160.

39. Jerphagnon, T; Pizzuti, M.G; Minnaard, A.J; Feringa, B.L. *Chem. Soc. Rev.,* **2009,** *38,* 1039-1075.

40. Holub, J.M; Kirshenbaum, K. *Chem. Soc. Rev.*, **2010**, *39*, 1325-1337.

41. Itagaki, M; Suenobu, K. *Org. Process Res. Dev.*, **2007**, *11*, 509–518

42. Agrawal, A; Kumari, S; Sahu, K.K. *Ind. Eng. Chem. Res.*, **2009**, *48,* 6145–6161

43. Itagaki, M; Masumoto, K; Suenobu, K; Yamamoto, Y. *Org. Process Res. Dev.,* **2006,** *10,* 245–250

44. Nelson, J. *Annu. Rep. Prog. Chem., Sect. A: Inorg. Chem.*, **2011**, *107*, 221-232.

45. Gamez, P; Aubel, P.G; Driessen, W.L; Reedijk, *J. Chem. Soc. Rev.*, **2001**, *30*, 376-385

46. Moga, T.G. *Nature Chemistry*, **2012,** *4*, 334

47. Barnett, S.M; Goldberg, K.I; Mayer, J.M. *Nature Chemistry*, **2012**, *4*, 498–502

48. Adzima, B.J; Tao, Y; Kloxin, C.J; DeForest, C.A; Anseth, K.S; Bowman, C.N. *Nature Chemistry*, **2011**, *3*, 256–259

49. Luo, S.G; Tan, H.M; Zhang, J.G; Wu, Y.J; Pei, F.K; Meng, X.H. *J Appl Polym Sci*, **1997**, *65*, 1217–1225

50. Gao, L.L; Liu, Y.H; Lei, H; Peng, H; Ruan, R. *Journal of Applied Polymer Science*, **2010** *116*, 1694–1699

51. Soest, P.J.V; Wine, R.H. *J Assoc Of Anal Chem*, **1968**, *51*, 780

52. Vogel's textbook of micro and semimicro qualitative inorganic analysis, 5[th] Ed., Longman Inc. publishers, The Chaucer Press, 1979.